\documentclass{ws-ijmpa}

\usepackage[super,compress]{cite}
\usepackage{hyperref}
\hypersetup{colorlinks,urlcolor=black,citecolor=black,linkcolor=black,filecolor=black}
\usepackage{breakurl}
\usepackage{epsfig}
\PassOptionsToPackage{hyphens}{url}
\usepackage[hyphens]{url}
\usepackage{subfigure}

\hypersetup{
  colorlinks=true,
  linkcolor=blue,
  citecolor=blue,
  urlcolor=blue
}

\begin{document}

\thispagestyle{plain}

\def\bib{B\kern-.05em{I}\kern-.025em{B}\kern-.08em}
\def\btex{B\kern-.05em{I}\kern-.025em{B}\kern-.08em\TeX}

\newcommand{\beq}{\begin{equation}}
\newcommand{\eeq}{\end{equation}}
\newcommand{\la}{\langle}
\newcommand{\promc}{{\sc ProMC}}
\newcommand{\ra}{\rangle}
\newcommand{\eps}{\epsilon}
\newcommand{\ud}{\mathrm{d}}
\newcommand{\Ec}{\mathcal{E}}
\newcommand{\Fc}{\mathcal{F}}
\newcommand{\Za}{\mathrm{Z_1}}
\newcommand{\Zb}{\mathrm{Z_2}}
\newcommand{\Zn}{\mathrm{Z_n}}
\newcommand{\F}{\mathrm{F}}

\chardef\til=126
\newcommand{\mev}{{\,\mathrm{MeV}}}
\newcommand{\gev}{{\,\mathrm{GeV}}}
\newcommand{\tev}{{\,\mathrm{TeV}}}

\markboth{S.Chekanov and M.~Demarteau}{Conceptual Design Studies for a CEPC Detector}

\catchline{}{}{}{}{}
%
%%%%%%%%%%%%%%%%%%%%%%%%%%%%%%%%%%%%%%%%%%%%%%%%%%%%%%%%%%%%%%%%%%%%

\title{
Conceptual Design Studies for a CEPC Detector 
}
%%%%%%%%%%%%%%%%%%%%%%%%%%%%%%%%%%%%%%%%%%%%%%%%%%%%%%%%%%%%%%%

%\author{S.V.~Chekanov\footnote{chekanov@anl.gov}\,  and M.~Demarteau\footnote{demarteau@anl.gov}}

\author{S.V.~Chekanov and M.~Demarteau}

\address{
HEP Division, Argonne National Laboratory,
9700 S.~Cass Avenue,
Argonne, IL 60439, USA. \\ 
Emails: chekanov@anl.gov, demarteau@anl.gov 
}

\maketitle

%\begin{history}
%\received{Day Month Year}
%\revised{Day Month Year}
%\end{history}

\begin{abstract}
The physics potential of the Circular Electron Positron Collider (CEPC) 
can be significantly strengthened by two detectors with
complementary designs. A promising detector approach  
based on the Silicon Detector (SiD) designed for the 
International Linear Collider (ILC) is presented. Several simplifications of this detector for the 
lower energies expected at the CEPC are proposed. A number of cost optimizations of this detector are illustrated 
using full detector simulations. We show that the proposed changes will enable to reach the physics goals at the CEPC. 

\end{abstract}

\ccode{PACS numbers: 13.66.-a, 13.66.Jn}

%%%%%%%%%%%%%%%%%%%%%%%%%%%%%%%%%%%%%%%%%%%%%%%%%%%%%%%%%%%%%%%%%%
\section{Introduction}
%%%%%%%%%%%%%%%%%%%%%%%%%%%%%%%%%%%%%%%%%%%%%%%%%%%%%%%%%%%%%%%%%%
  
The Circular Electron Positron Collider (CEPC) project 
is currently planned in China as a Higgs factory.
Operating at the center-of-mass (CM) energy of 250~GeV (or above), 
the CEPC experiment will take advantage of the clean environment of 
$e^+e^-$ collisions needed for high-precision measurements of the Higgs boson. CEPC experiments 
can  significantly strengthen
our understanding of the fundamental processes at the electroweak sector of the Standard Model (SM),
and can lead to discoveries of new physics through the precision measurements of the SM. 

In order to achieve the physics goals at the CEPC, a detector should be well optimized for 
measurements of $e^+e^-$ annihilation.
In particular, the studies of physics processes in the Higgs sector are 
considered to be the primary goal of the new experiment.
A promising approach for a detector at the CEPC  can be based on the Silicon Detector (SiD) 
concept \cite{Aihara:2009ad} developed for the International Linear Collider (ILC) \cite{Adolphsen:2013kya,Behnke:2013lya}. 
The design of this detector has a long history, and the experience gained during the R\&D phase of this
detector can be extremely valuable during the preparation to the CEPC concept. 

The abbreviation ``SiD'' stands for ``silicon detector'' --  
a compact general-multipurpose detector designed for high-precision  measurements of 
$e^+e^-$ annihilation at a CM energy of 500~GeV, which can be 
extended to 1~TeV. 
The choice of silicon for the tracking system and for the electromagnetic calorimeter ensures 
that the detector  
is robust to beam backgrounds, while a high-granular  calorimeter is well suited for the 
reconstruction of separate particles.
Some key characteristics of the SiD detector are: 

\begin{enumerate}

\item $4\pi$ solid angle coverage for reconstructed particles;  

\item Full 5-layer silicon tracking system with 50 $\mu m$  readout pitch size;

\item Silicon pixel detector with 20 $\mu m$  readout pitch size;

\item Superconducting solenoid with a 5 Tesla (T) field; 

\item Highly segmented silicon-tungsten electromagnetic calorimeter (ECAL) with the transverse cell size of 0.35~cm; 

\item Highly segmented hadronic calorimeter (HCAL) with a transverse cell size of $1\times 1$~cm. The depth of the HCAL
in the barrel region is about 4.5 interaction length\footnote{Nuclear interaction length, ($\lambda_I$),  is the average distance traveled by a hadronic particle before undergoing an inelastic nuclear interaction.}  ($\lambda_I$). The calorimeter has 40 longitudinal layers in the barrel and 45 layers in the endcap region; 

\end{enumerate}
Both ECAL and HCAL calorimeters are  finely segmented longitudinally and 
transversely. This is required for ``imaging'' capabilities of the calorimeter system:   
Together with the efficient tracking, the fine  segmentation of the calorimeter system optimizes the 
SiD detector for particle-flow algorithms (PFA)  
which allow identification and reconstruction of separate particles. 
The PFA objects can be reconstructed using the  	
Pandora Particle Flow algorithm  \cite{Charles:2009ta,Marshall:2013bda}.

The response of the SiD detector to physics processes
is simulated using the SLIC software package (''Simulator for the Linear Collider'')  \cite{Graf:2006ei}
developed for the ILC project.
The main strength of this software lies in the fact that  
it can easily be configured using XML option files, and it has
a platform-independent reconstruction which can be easily deployed on computers with different operating systems.  

The M\&S cost of the baseline design of the SiD  detector is estimated to be around \$320M \cite{Aihara:2009ad}, with 32\% being
allocated for the calorimeter, and 37\% for the magnet (estimated in 2009). 
% The ILC community has made multimillion investments 
% to the R\&D for this detector, including the development of the SLIC software used in 
% the past for performance studies by many  
% scientists.
 
\section{SiD for the CEPC }

For the CEPC physics goals, the SiD detector is over-designed. 
For example, the cost can be substantially reduced by simplifying the calorimeters and by 
reducing the magnetic field of the solenoid. 
Due to the lower CM energy of 250~GeV at the CEPC, a number of optimizations of the SiD detector 
are proposed:
 
\begin{enumerate}
\item
5~T solenoid field can be reduced to 4~T;  
\item
40 layers of HCAL can be reduced to 35 by removing 5 outer HCAL layers in the SiD design.
The remaining 35 layers of the steal absorber 
correspond to about 4.1 nuclear interaction length; 
\item
45 layers of the HCAL endcap can be reduced to 35 layers. This  makes the CEPC 
detector fully uniform from the point of view of the HCAL depth.
\end{enumerate}
The reason for the reduction of the solenoid field  lies in the fact that the typical track momentum measured at CEPC 
is a factor of two (four) smaller
compared to the 500 (1000)~GeV $e^+e^-$ collisions at the ILC. The magnetic field could be further reduced, 
but this will require a more 
detailed study than shown in this paper. 
Similarly, the reduction of the calorimeter depth is motivated by the fact that the maximum jet transverse 
momentum at the CEPC is 125~GeV, which is a factor two (four)  smaller than for the 500 (1000)~GeV $e^+e^-$  machine. 
In terms of the HCAL interaction length, the proposed 4 $\lambda_I$ calorimeter is similar to that of the OPAL experiment \cite{Ahmet:1990eg}.
The total absorber (steal and tungsten) of the ECAL and HCAL calorimeter
systems corresponds to about 5.1 $\lambda_I$.

In order to explore the possibility of optimization of  the SiD detector to a lower CM energy,
we use the HepSim Monte Carlo repository \cite{Chekanov:2014fga} with several benchmark processes for $e^+e^-$ collisions.
The $e^+e^-$ events at the 250~GeV CM energy were generated using the PYTHIA6 \cite{pythia} model. The following processes
were generated:

\begin{itemize}
\item
Fully inclusive QCD dijet process; 
\item
$Z$-boson production with the decays $Z\to e^+e^-$, $Z\to \mu^+\mu^-$, $Z\to \tau^+\tau^-$, $Z\to b\bar{b}$; 
\item
Higgs production ($Z^0 H$) with the decays $H \to 4 l$, $H \to \gamma\gamma$, $H \to \tau^+\tau^-$, $H \to b\bar{b}$.
The Higgs mass was set to 125~GeV. 
\end{itemize}
The events were simulated using the SiD detector geometry, and reconstructed using  
the SLIC package with Pandora PFA. 
The simulation and reconstruction steps were performed using the Open-Science Grid \cite{Pordes:2007zzb}.
Events before and after the simulation of the detector response were registered in the HepSim data catalogue. 

In the following, the original SiD detector geometry will be called SiDloi3. 
The number of reconstructed events after the SiDloi3 detector simulation and reconstruction was about ten thousand.
Most representative observables which are expected to be sensitive to the tracking and calorimeter performance of the SiD detector were analysed.
The obtained results (not shown) were found to be within 
the specification of the SiD detector described in Ref.~\cite{Behnke:2013lya}. 

The same data samples were simulated and reconstructed using the CEPC-optimized geometry 
discussed in the beginning of this section, i.e. with the solenoid field changed from 5~T to 4~T, and the HCAL calorimeter  depth
reduced from 40 (45) to 35 layers. 
In the following,  the SiD geometry after such modifications called SiDcc1. 
Full details of this detector geometry are available from the HepSim repository. 
To reduce computation time, the number of simulated and reconstructed events for the SiDcc1 detector 
were a factor two smaller than for the SiDloi3 simulation.

\begin{figure}
\centering
  \subfigure[] {
  \includegraphics[scale=0.45]{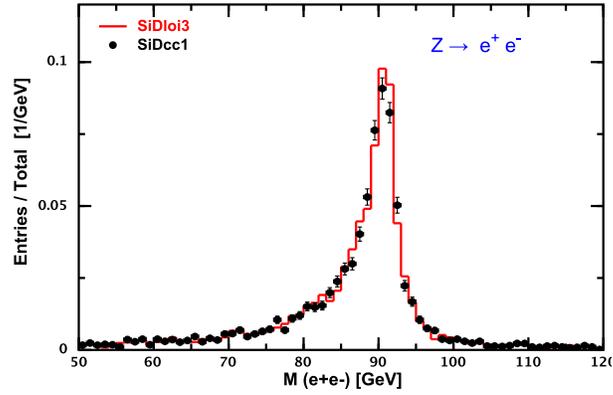}
  }
  \subfigure[] {
  \includegraphics[scale=0.45]{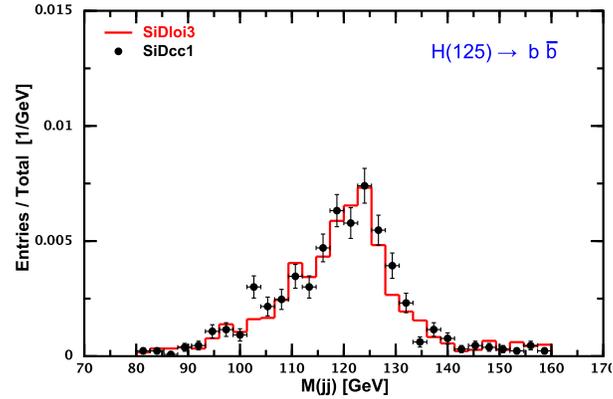}
  }
\caption{The invariant mass of two reconstructed electrons for the $Z\to e^+e^-$ process (a) and the invariant mass
of two jets for the process $H(125) \to b\bar{b}$ (b).  The distributions were reconstructed from the PFA objects.
The figure shows the original SiD setup (SiDloi3) and 
a CEPC optimized version of the SiD detector (SiDcc1). The distributions of the latter setup are shown as solid dots 
with statistical uncertainties.
}
\label{fig:masses}
\end{figure}

\begin{figure}
\centering
  \subfigure[] {
  \includegraphics[scale=0.45]{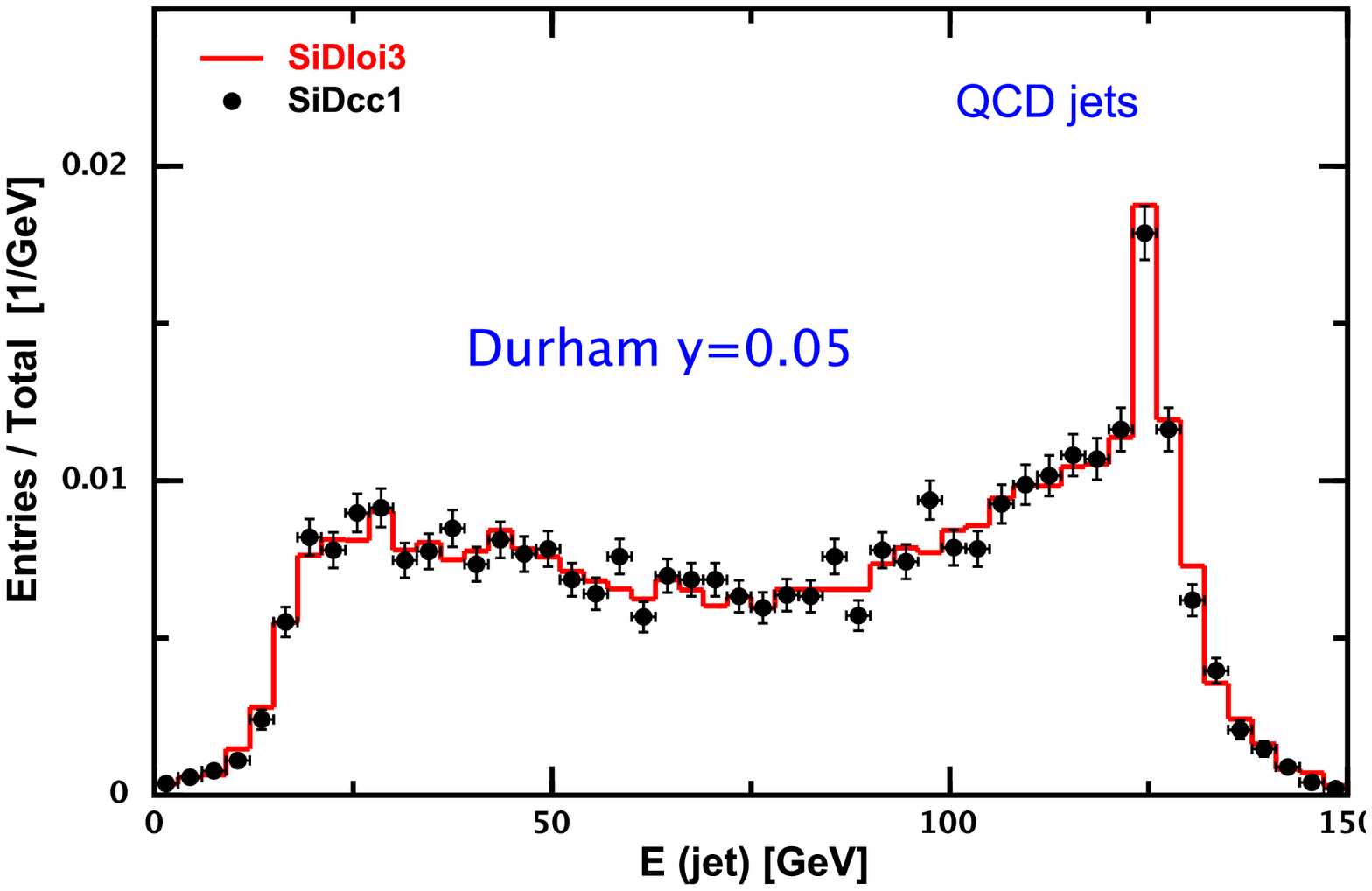}
  } 
  \subfigure[] {
  \includegraphics[scale=0.45]{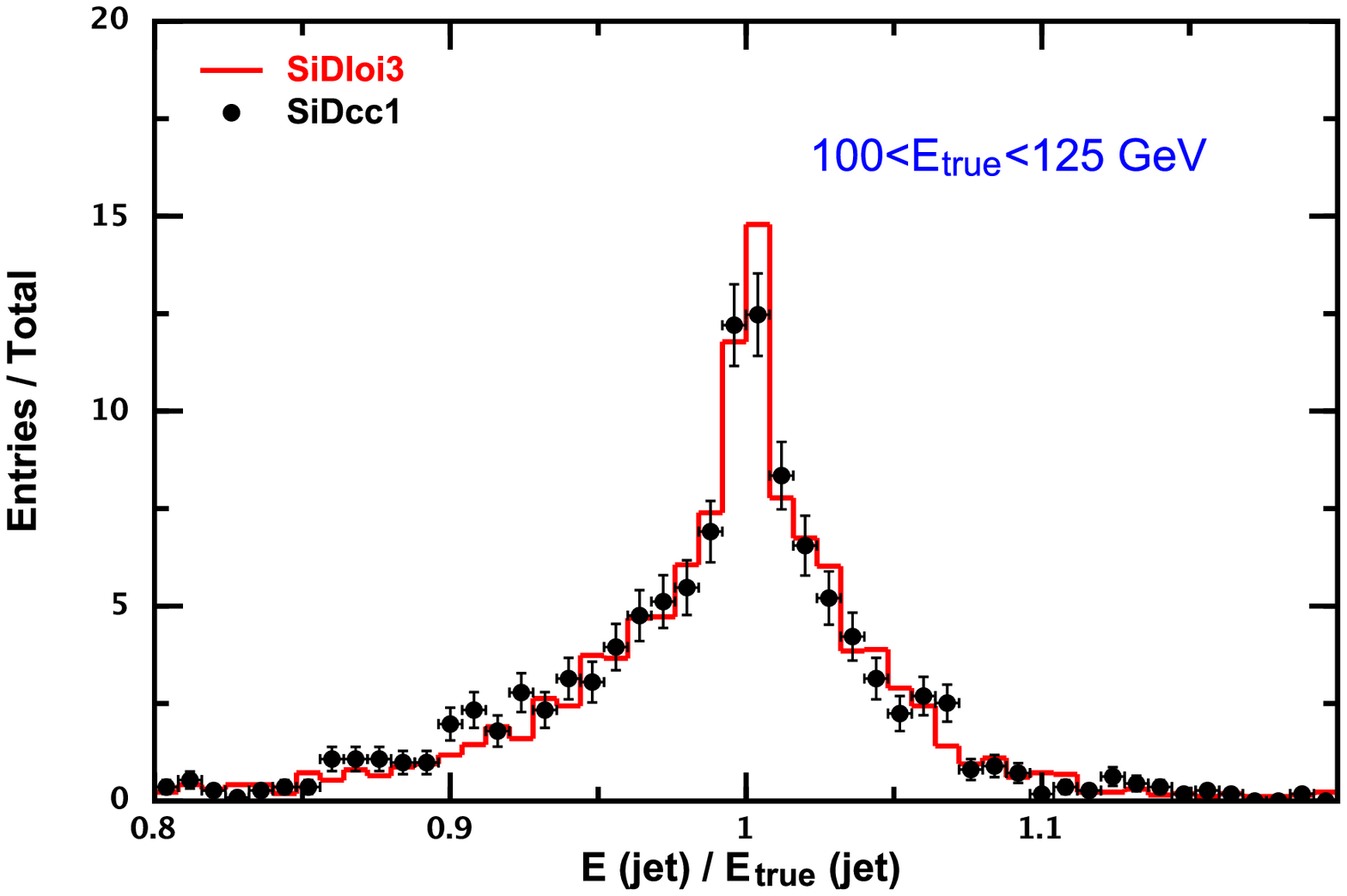}
  }
\caption{
The distribution of QCD jets in $e^+e^-$ collisions using the Durham algorithm with $y_{\mathrm{cut}}=0.05$ (a), and
the jet energy response for jets with energy close to the kinematic peak of 125~GeV.
The jets were reconstructed from the PFA objects.
The figure shows the original SiD setup (SiDloi3) and 
a CEPC optimized version of the SiD detector (SiDcc1). The distributions of the latter detector setup are shown as solid dots 
with statistical uncertainties.
}
\label{fig:jets}
\end{figure}

The distributions of several observables which are particularly sensitive to the change in the strength of 
the solenoid field and the HCAL absorber depth is shown in Figs.~\ref{fig:masses} and \ref{fig:jets}.
The distributions were reconstructed from the PFA objects which combine the information from four-momenta of tracks 
and calorimeter energy deposits.
For example, the  $Z$ boson mass reconstructed from the invariant mass
of two electrons (Fig.~\ref{fig:masses}(a)) is sensitive 
to the performance of tracking system to high-momentum tracks ($e^+/e^-$).
The energy distribution of hadronic jets reconstructed from the PFA objects is sensitive to both the performance of the 
tracking system,  and to the HCAL longitudinal segmentation.
Figure~\ref{fig:masses}(b) shows the invariant mass of two jets for the process $H(125) \to b\bar{b}$.
The jets were reconstructed with the Jade algorithm \cite{jade}  by forcing two jets per event, and requiring the
transverse momentum of jets to be above $20$~GeV.

To take a closer look at the hadronic jets, 
Figure~\ref{fig:jets} shows the distribution of the jet transverse momentum for inclusive QCD processes in $e^+e^-$ at 250 GeV. The jets 
were reconstructed using the Durham algorithm \cite{durham} 
with the parameter $y_{\mathrm{cut}}=0.05$. As before, the input to this algorithm are the PFA objects.
Jets were selected with a minimum transverse momentum of $20$~GeV. 
Figure~\ref{fig:jets}(b) shows the jet energy response by taking the ratio of the reconstructed jet energy to the energy of jets reconstructed  
from stable particles, which are  
defined if their lifetime $\tau$ are smaller than $3\cdot 10^{-10}$ seconds. Neutrinos were excluded from consideration.
As expected, the distributions for this ratio peaks at one, indicating
that no energy leakage is observed for both the SiDloi3 and SiDcc1 detectors.

\begin{figure}
\centering
  \includegraphics[scale=0.5]{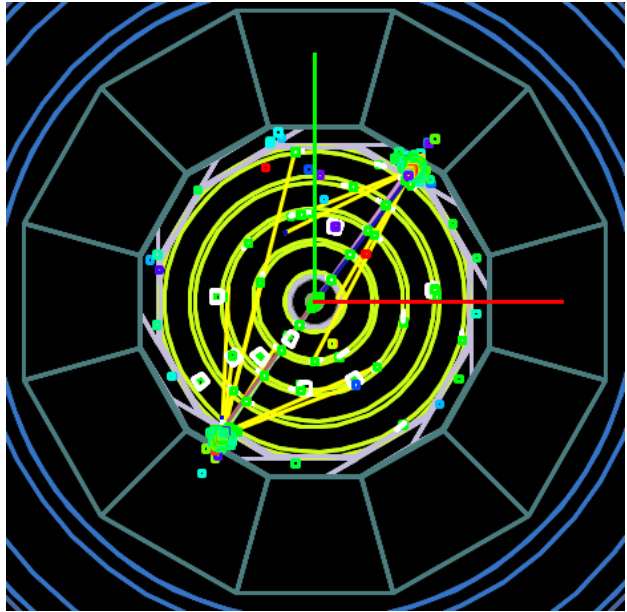}
\caption{An event display of an $Z\to e^+e^-$ event using the optimized SiDcc1 setup. 
Clusters of the green points on the surface of the ECAL correspond to the reconstructed $e^+/e^-$ from the $Z$ decay. 
The space between the outer layer of HCAL and the solenoid 
is due to removed 5 outer HCAL layers of the original SiD detector.
}
\label{fig:zee}
\end{figure}

Figure~\ref{fig:zee} illustrates a typical $Z\to e^+e^-$ event in the Jas3/Wired4 event display. 
A prominent feature of this event is the energy deposits in the 
ECAL corresponding to the electrons from the $Z$ decay.  The space between the outer layer of the HCAL and the solenoid  
is due to the removal of 5 HCAL layers from the original design of the SiD detector.

In summary, this paper suggests that 
the SiD detector (or its sub-detectors)  can be re-purposed for the CEPC.
We have illustrated a few directions to optimize the SiD detector for lower CM energies.
The results obtained with the SiDloi3 and SiDcc1 detector concepts show good agreements (within statistical errors), thus
the optimized SiDcc1 detector will enable to reach the physics goals at the CEPC.
It should be noted that the changes to the SiD concept listed  above are just a 
few possible options to reduce the cost of a detector for the CEPC energy,  without compromising the physics goals at the CEPC.
It is very likely that a more substantial optimization can be made after dedicated performance  studies.

\section*{Acknowledgments}

This research was done using resources provided by the Open Science Grid,
which is supported by the National Science Foundation and the U.S. Department of Energy's Office of Science. 

The submitted manuscript has been created by UChicago Argonne, 
LLC, Operator of Argonne National Laboratory (Argonne). Argonne, a U.S. Department of Energy Office of Science laboratory, 
is operated under Contract No. DE-AC02-06CH11357. Argonne National Laboratory's work was supported by the U.S. 
Department of Energy under contract DE-AC02-06CH11357. 
The U.S. Government retains for itself, and others acting on its behalf, a paid-up nonexclusive, 
irrevocable worldwide license in said article to reproduce, prepare derivative works, distribute copies to 
the public, and perform publicly and display publicly, by or on behalf of the Government.

%%%%%%%%%%%%%%%%%%%%%% references %%%%%%%%%%%%%%%%%%%%%%%%%%%%%%
\bibliographystyle{ws-ijmpa}
\bibliography{biblio}

\end{document}